\documentclass[11pt,a4paper]{article}
\usepackage[english]{babel}
\usepackage[OT1]{fontenc}
\usepackage{hyperref}

\makeatletter
\def\section{\@startsection{section}{1}{\z@}{-3.5ex plus -1ex minus -2.ex}
{2.3ex plus .2ex}{\Large\bf}}

\def\subsection{\@startsection{subsection}{2}{\z@}{-3.25ex plus
 -1ex minus -2.ex}
{1.5ex plus .2ex}{\bf}}

\def\f{\newline}
\def\e{\varepsilon}
\def\comp{{\rm comp}}
\def\rr{{\bf R}}

\newcommand{\be} {\begin{equation}}
\newcommand{\ee} {\end{equation}}
\newcommand{\bd} {\begin{displaymath}}
\newcommand{\ed} {\end{displaymath}}
\newcommand{\bq}{\begin{eqnarray}}
\newcommand{\eq}{\end{eqnarray}}
\newcommand{\bqn}{\begin{eqnarray*}}
\newcommand{\eqn}{\end{eqnarray*}}
\newcommand{\ba}[1]{\begin{array}{#1}}
\newcommand{\eqa}{\end{array}}
\def\qed{
   \\[-4ex]
  \hbox to \hsize{\hfill \vrule height 1.6ex width 1.5ex
  depth -.1ex}}

\newenvironment{proof}[1][Proof]{\ \\ {\bf #1:}\par \noindent}{\qed\\}

\newcommand{\eeq}{e^{{\rm quant}}}
\newcommand{\eew}{e^{{\rm worst}}}
\newcommand{\eer}{e^{{\rm rand}}}
\newcommand{\ccq}{\comp^{{\rm quant}}}
\newcommand{\ccw}{\comp^{{\rm worst}}}
\newcommand{\ccr}{\comp^{{\rm rand}}}

\newcommand{\A}{A^*}

\newtheorem{theorem}{Theorem}
\newcommand{\cost}{{\rm cost}}

\begin{document}
\begin{center}\LARGE \bf{On the Complexity of Searching Maximum of a Function on a Quantum
Computer}\end{center}
\medskip
\begin{center}
Maciej Go\'cwin\footnotemark[1]
\end{center}
\footnotetext[1]{\begin{minipage}[t]{16cm}
 \noindent
{\it Department of Applied Mathematics, AGH University of Science
and Technology,\\
\noindent  Al. Mickiewicza 30, paw. B7, II p.,
pok. 24,\\
 30-059 Cracow, Poland
\newline
 gocwin@uci.agh.edu.pl, tel. +48(12)617 4405 }
\end{minipage} }
\thispagestyle{empty}
\begin{abstract}
\noindent We deal with a problem of finding maximum of a function
from the H\"older class on a quantum computer. We show matching
lower and upper bounds on the complexity of this problem. We prove
upper bounds by constructing an algorithm that uses the algorithm
for finding maximum of a discrete sequence. To prove lower bounds
we use results for finding logical OR of sequence of bits. We show
that quantum computation yields a quadratic speed-up over
deterministic and randomized algorithms.
\end{abstract}
\begin{section}{Introduction}\label{sec1} Quantum algorithms yield
a speed-up over deterministic and Monte Carlo algorithms for many
problems. Many papers deal with quantum solution of discrete
problems, starting from the work of Shor \cite{shor}, followed by
database search algorithm of Grover \cite{grover}. Other discrete
problems were also studied, such as discrete summation,
computation of the mean, median and $k$th-smallest element
\cite{brassard},\cite{brassard2},\cite{durr},\cite{grover2},\cite{nayak}.
 \\
There is also a progress in studying the quantum complexity of
numerical problems. The first paper dealing with a continuous
problem was that of Novak \cite{novak1}, who considered
integration of a function from the H\"older class. The problem of
function approximation on quantum computer was studied by Heinrich
\cite{heinrich1},\cite{heinrich2}. Also path integration
\cite{wozniakowski} and differential equations
\cite{kacewicz1},\cite{kacewicz2} on a quantum computer were
investigated.
\\
In this paper, we deal with a problem of finding maximum of a
function from the H\"older class on a quantum computer. The
complexity of this problem
in deterministic and randomized settings on a classical computer is well known \cite{novak}.\\
We present matching upper and lower complexity bounds in the
quantum setting. We show that quantum computations yield a
quadratic speed-up compared to deterministic and randomized
algorithms over entire range of class parameters. Upper bounds are
shown by constructing a suitable algorithm, which uses the optimal
algorithm for finding maximum of a discrete sequence. To prove the
lower bound we use the result of Nayak and Wu \cite{nayak} for
finding logical OR of sequence of bits.
\\
In the next section necessary definitions are presented. Existing
results for searching maximum of a sequence are shown in Section
\ref{sec3}. The main result of this paper is contained in Theorem
\ref{th1} in Section \ref{sec4}.
\end{section}
\begin{section}{Quantum setting}\label{sec2}
In this section we briefly describe the model of computation. More
details about quantum computing can be found in \cite{novak1}.
Suppose that we have a numerical problem given by a solution
operator $S:F\rightarrow G$, where $F$ is a subset of a linear
function space and $G$ is a normed space. We are interested in
computing an approximation of $S(f)$ for $f\in F$ on a quantum
computer. This is done by an algorithm $A$. The algorithm can
access the input element $f$  only by a quantum query. An output
of the algorithm $A$ for a given $f$ is a random variable
$A(f,\omega)$. For a detailed discussion of quantum algorithms and
the quantum query operator, the reader is referred to
\cite{novak1}.
\\
We know recall what is meant by the error of an algorithm. Let
$0<\theta<1$. The local error of algorithm $A$ on input element
$f$ is defined by \bd
\eeq(S,A,f,\theta)=\inf\{\e:P\{\|S(f)-A(f,\omega)\|>\e\}\leq\theta\}.\ed
The number $1-\theta$ is thus the lower bound on success
probability. For $\e>0$ the bound $\eeq(S,A,f,\theta)\leq\e$ holds
iff algorithm $A$ computes $S(f)$ with error at most $\e$ and
probability at least $1-\theta$. The global error in the class $F$
is defined as \bd \eeq(S,A,F,\theta)=\sup_{f\in
F}\eeq(S,A,f,\theta).\ed For $\theta=1/4$, we denote the error for
$f$ by $\eeq(S,A,f)$ and the error in the class $F$ by
$\eeq(S,A,F)$.\\ The cost of the algorithm on the input element
$f$,  $\cost(A,f)$ is defined as a number of accesses to an
oracle. In classical settings it is a number of function values or
derivative values that is used to compute maximum of the function.
In the quantum setting, by an oracle we mean the quantum query
operator. The global cost of algorithm is defined as \bd
\cost(A,F)=\sup_{f\in F}\cost(A,f).\ed For $\e>0$, the
$\e$-complexity is defined as the minimal cost of an algorithm
that produces an $\e$-approximation: \bd
\ccq_{\e}(S,F)=\min_A\{\cost(A,F)|\ \eeq(S,A,F)\leq\e\}. \ed
\\
In the next section some known results about complexity of finding
maximum of a sequence of numbers in the quantum and classical
settings are presented.
\end{section}
\begin{section}{Searching  maximum of a sequence}\label{sec3}
We recall results on a discrete maximization (minimization)
problem. Consider the following problem: given a sequence
$X=(x_0,x_1,\ldots,x_{n-1})$ of real numbers in $[0,1]$, find the
number $x_i=\max(\min)\{x_j:j=0,1,\ldots ,n-1\}$. The cost of an
algorithm is defined as a number of accesses to the oracle, which
returns the input number $x_i$, $i=0,1,\ldots,n-1$. Another
possibility is to count a number of comparisons. In this model the
oracle returns the logical value of a comparison $x_i<x_j$,
where $i,j\in\{0,1,\ldots,n-1\}$.\\
Clearly, the complexity of this problem on a classical computer is
\be\ccw(n)=\Theta(n)\quad,\quad \ccr(n)=\Theta(n),\label{worst}\ee
where by  $\comp(n)$ we mean the minimal cost of an algorithm
computing the maximum (minimum) value from the sequence of $n$
numbers in suitable setting. Better results can be obtained on a
quantum computer. In 1996 C. D\"urr and P. H\o yer in \cite{durr}
presented comparison quantum algorithm for finding the minimum.
This algorithm finds minimum value from the list of $n$ items with
probability greater than ${1}/{2}$ and its running time is
$O(\sqrt{n})$. They based their algorithm on quantum exponential
searching algorithm \cite{boyer}, which is a generalization of
Grover's search algorithm introduced in \cite{grover}. This result
establishes the upper bound on the complexity of the problem of
finding maximum of a discrete
sequence on a quantum computer.\\
Lower bounds on this problem were established by A. Nayak and F.
Wu in \cite{nayak}. They examine the more general problem: for
$X=(x_0,,x_1,\ldots,x_{n-1})\in [0,1]^n$ and $\Delta>0$ compute
$\Delta$-approximate $k$th-smallest element, i.e., a number $x_i$
that is a $j$th-smallest element of $X$ for some integer
$j\in(k-\Delta,k+\Delta)$.\\
If $\Delta=1$ (or less) this problem reduces to the problem of
finding $k$th-smallest element exactly. For
$k=n-1$ $k$th-smallest element is the maximum value from the sequence.\\
In \cite{nayak}, a quantum algorithm has been presented with the
cost $O(N\log(N)\log\log(N))$, where
$N=\sqrt{n/\Delta}+\sqrt{k(n-k)}/\Delta$. The algorithm is
inspired by the minimum finding algorithm of D\"urr and H\o yer
\cite{durr}, and uses exponential search algorithm of Boyer
\emph{et al.} \cite{boyer}. It finds $\Delta$-approximation of
$k$th-smallest element for any $k\in\{0,\ldots,n-1\}$ and
$\Delta\geq1$ with probability at leat $2/3$.\\
Nayak and Wu in \cite{nayak} established essentially matching
lower bounds for this problem. To derive the bounds they used
polynomial method introduced by R. Beals \emph{et al.} in
\cite{beals}.\\ These results show that the complexity of
searching maximum of $n$ elements on a quantum computer is of
order \be\ccq(n)=\Theta(\sqrt{n})\label{comp}.\ee The comparison
of this result with (\ref{worst}) shows that quantum computers
make a quadratic speed-up over classical computers for this
problem.
\end{section}
\begin{section}{Searching maximum of a function}\label{sec4}
We consider the problem of finding the maximum of a function from
the H\"{o}lder class \bqn
\lefteqn{F^{r,\rho}_{d}=\left\{f:[0,1]^d\rightarrow \rr \ |\ \
f\in C^r,\
\|f\|\leq1,\right.} \\
&&\left.\left|D^{(r)}f(x)-D^{(r)}f(y)\right|\leq\|x-y\|^\rho \ \
\forall x, y\in[0,1]^d,\ \ \forall D^{(r)}\right\}, \eqn where
$D^{(r)}$ run through the set of all partial derivatives of order
$r$, $r\in \bf{ N}_0$, $0<\rho\leq1$ and
$\|\cdot\|=\|\cdot\|_{\infty}$. We want to find a number \bd
M(f)=\max_{t\in[0,1]^{d}}f(t)\ed up to some given precision
$\e>0$, for any function $f$ from class $F^{r,\rho}_{d}$ with
probability not less than $3/4$.
\\
The complexity of this problem on a classical computer in the
deterministic
and randomized settings is presented in \cite{novak}. We shall recall these results for a further comparison.\\
\\
In the deterministic worst-case setting the local error of
algorithm $A$ on input function $f\in F_d^{r,\rho}$ is defined by
\bd \eew(M,A,f)=|M(f)-A(f)|\ed and the global error by \bd
\eew(M,A,F_d^{r,\rho})=\sup_{f\in F_d^{r,\rho}}\eew(M,A,f).\ed In
the randomized setting, algorithm
$A=(A(\omega))_{\omega\in\Omega}$ is a random variable on some
probabilistic space $(\Omega,B,m)$. The local error of this
algorithm is defined by \bd
\eer(M,A,f)=\int_{\Omega}|M(f)-A(\omega)(f)|d m(\omega),\ed and
the global error by \bd \eer(M,A,F_d^{r,\rho})=\sup_{f\in
F_d^{r,\rho}}\eer(M,A,f).\ed The cost of an algorithm in the
deterministic and randomized settings is meant as a number of
function values accessed by an algorithm. In the randomized
setting, points where $f$ is evaluated can be chosen randomly.\\
It is shown in \cite{novak} (pp. 34 and 59) that the complexity of
function maximization in the H\"older class in both worst-case and
randomized settings is given by \bd
\ccw_{\e}(M,F_d^{r,\rho})=\Theta\left(\left(\frac{1}{\e}\right)^{\frac{d}{r+\rho}}\right)\quad\textrm{and}\quad\ccr_{\e}(M,F_d^{r,\rho})=\Theta\left(\left(\frac{1}{\e}\right)^{\frac{d}{r+\rho}}\right)
.\ed
\\
We now pass to the quantum setting. In the following result we
prove that a significant improvement is achieved on a quantum
computer.
\begin{theorem}\label{th1}Let $\e>0$. The quantum $\e$-complexity of computing maximum of a function from H\"{o}lder
class $F^{r,\rho}_{d}$ is
\bd\ccq_{\e}(M,F_d^{r,\rho})=\Theta\left(\e^{-\frac{d}{2(r+\rho)}}\right).\ed
\end{theorem}
\newpage
\begin{proof}
First we prove the upper bound. We divide each edge of the cube
$[0,1]^d$ into  $n$ subintervals of equal length. We get $N=n^d$
cubes $K^i$, $i=1,\ldots ,N$. Let $t^i$ denote the center of cube
$K^i$. On every cube $K^i$ we use Taylor's expansion of $f$. For
$t\in K^i$ we have \bd f(t)=w^i(t)+R_r(t,t^i),\ed where \bd
w^i(t)=\sum_{k=0}^r \frac{1}{k!} f^{(k)}(t^i) (t-t^i)^k,\ed and
\bd R_r(t,t^i)=\int_0^1\left( f^{(r)} (\theta
t+(1-\theta)t^i)-f^{(r)}(t^i)\right) (t-t^i)^r
\frac{(1-\theta)^{r-1}}{(r-1)!}{\rm d}\theta.\ed Let \bd
m_i(f)=\max_{t\in K^i}w^i(t).\ed We consider the algorithm $\A$ of
the form \bd \A(f)=\max_{i=1,\ldots,N}\widetilde{m}_i(f),\ed where
$\widetilde{m}_i(f)$ is an approximation of $m_i(f)$ computed by
some classical algorithm. We assume that \bd
|\widetilde{m}_i(f)-m_i(f)|\leq\e_1 \quad \forall i=1,\ldots,N,\ed
for some $\e_1>0$ independent of $i$ and $f$. To compute
$\widetilde{m}_i(f)$ on a classical computer we do not need any
new evaluations of $f$ or its partial derivatives, so that
information cost does not increase. (Of course, it is still not an
easy task to compute $\widetilde{m}_i(f)$ and it increases
combinatory cost of the algorithm.) The maximum of the discrete
set of numbers $\widetilde{m}_1(f),\ldots,\widetilde{m}_N(f)$ we
compute on a quantum computer, with probability not less than
$3\over4$. This is done by the optimal algorithm described in
Section \ref{sec3}.\f  We now estimate the error of the algorithm
defined above \bq \label{eq1}
\eeq(M,\A,f)&=&|M(f)-\A(f)|=|\max_{t\in[0,1]^d}f(t)-\max_{i=1,\ldots ,N } \widetilde{m}_i(f)|\nonumber\\
&=&|\max_{i=1,\ldots ,N }\max_{t\in K^i}f(t)-\max_{i=1,\ldots ,N }
\widetilde{m}_i(f)|\nonumber\\
&\leq& \max_{i=1,\ldots ,N }|\max_{t\in
K^i}f(t)-\widetilde{m}_i(f)|\nonumber\\
&\leq& \max_{i=1,\ldots ,N }\left(|\max_{t\in
K^i}f(t)-m_i(f)|+|m_i(f)-\widetilde{m}_i(f)|\right)\nonumber\\
&=& \max_{i=1,\ldots ,N }|\max_{t\in K^i}f(t)-\max_{t\in
K^i}w^i(t)|+\e_1\nonumber\\
&\leq&\max_{i=1,\ldots ,N }\max_{t\in K^i}|f(t)-w^i(t)|+\e_1 .\eq
For $t\in K^i$ we have
\bqn |f(t)-w^i(t)|&=&|R_r(t,t^i)|\\
&=&\left|\int_0^1\left( f^{(r)} (\theta
t+(1-\theta)t^i)-f^{(r)}(t^i)\right) (t-t^i)^r
\frac{(1-\theta)^{r-1}}{(r-1)!}{\rm d}\theta\right|\\
&\leq&\sup_{\theta\in[0,1]}|f^{(r)} (\theta
t+(1-\theta)t^i)-f^{(r)}(t^i)|\;\|t-t^i\|^r \int_0^1\frac{(1-\theta)^{r-1}}{(r-1)!}{\rm d}\theta\\
&\leq&H\sup_{\theta\in[0,1]} \|\theta
t+(1-\theta)t^i-t^i\|^{\rho}\;\|t-t^i\|^r\\
&=&H\; \|t-t^i\|^{r+\rho}\leq H \left(\frac{1}{n}\right)^{r+\rho}.
\eqn The constant $H$ depends on $d$ and $r$ but not on $n$. From
this, and inequality~(\ref{eq1}) we get the error bound \bd
\eeq(M,\A,f)\leq H \left(\frac{1}{n}\right)^{r+\rho}+\e_1.\ed We
now choose $\e_1=(1/n)^{r+\rho}$. Then for some constant $G$
independent of $n$ we have \be\label{eq2} \eeq(M,\A,f)\leq
G\left(\frac{1}{n}\right)^{r+\rho}.\ee We examine the cost of
algorithm $\A$.\f To compute $\widetilde{m}_i(f)$ we need to know
the value of $f$  and the values of all its partial derivatives of
order up to $r$ at point $t^i$. So the cost of computing
$\widetilde{m}_i(f)$ is \bd \cost(\widetilde{m}_i)=\sum_{k=0}^r
\left({d+k-1}\atop{k}\right)=\frac{(d+r)!}{d!\,r!},\ed which is
independent on $n$.\f Due  to (\ref{comp}) the cost of computing
the  maximum of numbers
$\widetilde{m}_1(f),\ldots,\widetilde{m}_1(f)$ on quantum computer
is $O(\sqrt{N})=O(\sqrt{n^d})$ accesses to the numbers
$\widetilde{m}_i(f)$. Thus, the total cost is \be\cost(\A,f)\leq
C\; n^{d/2}\ee for some constant $C$. Due to (\ref{eq2}), to
obtain $\eeq(M,\A,f)\leq \e$ it suffices to take
$K\;\e^{-\frac{d}{2(r+\rho)}}$ function and derivative values,
where $K$ is a constant. This completes the proof of the upper
bound.
\\
\\
We now prove the lower bound. Assume that $A$ is any algorithm
that computes $M(f)=\max_{t\in [0,1]^d}f(t)$ for any $f\in
F^{r,\rho}_d$ up to the error $\e$ with probability not less than
$3\over4$. We denote the cost of this algorithm by $c(\e)$.\f For
$\e_1>0$, the class $F^{r,\rho}_d$ contains
$n=\Theta\left(\e_1^{-\frac{d}{r+\rho}}\right)$ functions
$f_1,\ldots,f_n$ with disjoint supports such that
$\max_{t\in[0,1]^d}f_i(t)=\e_1$ (see \cite{novak}, p. 35).\f Let
$\e_1=4\e$. Let $X=(x_1,\ldots,x_n)$ be any sequence such that
$x_i\in\{0,1\}\quad\forall i=1,\ldots,n$. Then the function \bd
f_{\e_1}:=\sum_{i=1}^n x_i f_i\ed belongs to the class
$F_d^{r,\rho}$. Thus, algorithm $A$ applied to $f_{\e_1}$ computes
$\max_{t\in[0,1]^d}f_{\e_1}(f)$ up to the error
$\e=\frac{\e_1}{4}$, with the cost $c(\e)=c(\frac{\e_1}{4})$. That
is \be\label{eq3}
|\max_{t\in[0,1]^d}f_{\e_1}(t)-A(f_{\e_1})|\leq\frac{\e_1}{4} \ee
with probability not less than $3\over4$, and cost
$c(\e)=c(\frac{\e_1}{4})$.\\
 From the
definition of $f_{\e_1}$ we see that \be
\label{eq4}\setlength\arraycolsep{2ex}
\max_{t\in[0,1]^d}f_{\e_1}(t)=\left\{\begin{array}{l l l}
\e_1&\,\textrm{if}&\max_{i=1,\ldots,n}x_i=1 \\
0&\,\textrm{if}&\max_{i=1,\ldots,n}x_i=0 \end{array}\right..\ee If
$A(f_{\e_1})\geq {3\over 4} $, then due to (\ref{eq3}) \bd
\max_{t\in[0,1]^d}f_{\e_1}(t)\geq A(f_{\e_1})-{1\over 4} \e_1\geq
{1\over 2} \e_1. \ed So, in this case, due to (\ref{eq4}),
$\max_{t\in[0,1]^d}f_{\e_1}(t)=\e_1$ and
$\max_{i=1,\ldots,n}x_i=1$ with probability at least $3\over4$.
Similarly, with probability at least $3\over4$, if
$A(f_{\e_1})\leq {1\over 4} \e_1 $, then
$\max_{t\in[0,1]^d}f_{\e_1}(t)=0$ and $\max_{i=1,\ldots,n}x_i=0$.
\\
Based on algorithm $A$, we now define an algorithm $\tilde{A}$,
which finds the maximum of a sequence $X=(x_1,\ldots,x_n)$. This
algorithm is constructed as follows: \bqn \textrm{if }
&{3\over4}\e_1\leq A(f_{\e_1})\leq
{5\over4}\e_1&\textrm{, then we put }\tilde{A}(X)=1\textrm{, and}\\
\textrm{if }&-{1\over4}\e_1\leq
A(f_{\e_1})\leq{1\over4}\e_1&\textrm{, then we put
}\tilde{A}(X)=0.\eqn In the other cases we put $\tilde{A}(X)=0$.
With probability at least $3\over4$, we have that
$\max_{i=1,\ldots n}x_i=0$ and $\tilde{A}(X)=0$, or
$\max_{i=1,\ldots n}x_i=1$ and $\tilde{A}(X)=1$. Hence, the
algorithm $\tilde{A}$ computes the maximum of $n$ numbers
$x_1,\ldots,x_n$, such that $x_i\in\{0,1\}$ (logical OR of the
input bits), with probability not less than $3\over4$. From
\cite{nayak} (see proof of Theorem 1.5 in the case of
$x_i\in\{0,1\}$) we know that the cost of such an algorithm is
$\Omega(\sqrt{n})$. The cost of the algorithm $A$ is not less than
the cost of the algorithm $\tilde{A}$. Since \bd
n=\Theta\left(\left({1\over\e_1}\right)^{\frac{d}{r+\rho}}\right)=\Theta\left(\left({1\over\e}\right)^{\frac{d}{r+\rho}}\right),\ed
 we have that \bd
c(\e)=\Omega\left(\sqrt{n}\right)=\Omega\left(\left({1\over\e}\right)^{\frac{d}{2(r+\rho)}}\right).
\ed This completes the proof of the theorem.
\end{proof}
\\ \\
Comparing this to the classical deterministic or random complexity
of this problem, which is
$\Theta\left(\e^{-\frac{d}{r+\rho}}\right)$, we see that a quantum
computer makes a quadratic speed-up over classical settings. This
is achieved over the entire range of $r$, $\rho$ and $d$. For the
integration problem, a quadratic speed-up over the randomized
setting holds only for  $(r+\rho)/d$ small.

\end{section}

\end{document}